\documentclass[12pt]{article} 
\usepackage{amsmath}
\usepackage{amssymb} 
\usepackage{amscd}
\usepackage{amsthm}
\usepackage[utf8]{inputenc}
\usepackage{hyperref}
\def\gr{\mathrm{gr}}

\def\det{{\mbox{det}}}

 \def\fraci{{\mathfrak I}}

\def\Fg{{\mathbf F}}

\numberwithin{equation}{section}

\begin{document}
\enlargethispage{3cm}

\thispagestyle{empty}
\begin{center}
{\bf ABUNDANCE OF LOCAL ACTIONS}
\end{center} 
\begin{center}
{\bf  FOR THE VACUUM EINSTEIN EQUATIONS}
\end{center}
   
\vspace{0.3cm}

\begin{center}
Michel DUBOIS-VIOLETTE
\footnote{Laboratoire de Physique Th\'eorique, UMR 8627\\
Universit\'e Paris XI,
B\^atiment 210\\ F-91 405 Orsay Cedex\\
Michel.Dubois-Violette$@$u-psud.fr} and 
Mohamed LAGRAA\footnote{Laboratoire de Physique Théorique,\\ Universit\'e d'Oran Es-Senia,\\ 31100 Alg\'erie\\ m.lagraa@lycos.com}
\end{center}
\vspace{0,5cm}
 \vspace{0,5cm}

\begin{abstract}
We exhibit large classes of local actions for the vacuum Einstein equations. In presence of fermions, or more generally of matter which couple to the connection, these actions lead to inequivalent equations revealing an arbitrary number of parameters. Even in the pure gravitational sector, any corresponding quantum theory would depend on these parameters.
\end{abstract}
\vfill

\noindent LPT-ORSAY 09/55
\newpage
\section{Introduction}
The addition of the Holst term to the ``tetrad-connection" form of the Einstein gravitational action has attracted much attention these last few years in connection with the development of nonperturbative methods of quantization of gravity \cite{rov:2007}, \cite{thi:2008} (and references therein) which have followed the emergence of new canonical variables in general relativity  \cite{ash:1987},   \cite{bar:1995}. The corresponding modified action which is still local depends on a new dimensionless parameter $\gamma$ refered to as the {\sl Immirzi parameter}. As long as one considers classical gravity without sources this addition is immaterial (whenever $\gamma^2+1\not= 0$ for the Lorentzian case) in the sense that it leads to the classical vacuum Einstein equations. However as soon as one adds fields which couple to the connection the equations are modified and do depend on $\gamma$ \cite{per-rov:2006} (note however that the minimal coupling procedure is somehow ambiguous \cite{Kaz:2008}). Furthermore even without addition of such fields, for any quantization procedure the corresponding quantum theory does depend on $\gamma$, see e.g.in \cite{rov:2007}. Although there are very good reasons to introduce the Holst action, this raises the following natural questions. Are there many non trivial local actions which lead to the classical vacuum Einstein equations ? Our answer is yes (in any dimension) and the object of this article is to exhibit an infinity of such actions.\\
In order to precise our notations and to make clear the origin of the arbitrariness of the actions, we first review in the next section the Einstein-Cartan formalism and give different descriptions of the Holst term. Then in Section 3, we discuss addition of torsion terms to the Einstein action in arbitrary dimension $s+1$. In Section 4 we discuss the addition of the same type of terms to the complete 4-dimensional Holst action. Section 5 is our conclusion.\\
Throughout this paper we discuss the Lorentzian case in dimension $s+1$ with $s\geq 2$ but it is clear that similar constructions work for the other signatures in particular in the Riemannian case.

\section{The Einstein-Cartan formalism}

In the Einstein-Cartan formalism the variables describing the gravitation are a local coframe $\theta^a$ ($a\in \{0,\dots,s\}$) and a linear connection which is metric for the metric $g$ for which the $\theta^a$ are orthonormal i.e. $g=\eta_{ab}\theta^a\otimes \theta^b$ where the $\eta_{ab}$ are the components of the flat metric diag ($-+ \dots +$). Using $\eta$ and its inverse to lower and to lift the (Lorentz) indices $a,b,c \dots$, the connection is given by local 1-forms $\omega^{ab}=-\omega^{ba}$. The structure equations induce then
\begin{equation}
d\theta^a+\omega^{a\phantom{a}}_{\phantom{a}b}\wedge \theta^b=\Theta^a
\label{StT}
\end{equation}
\begin{equation}
d\omega^{ab}+\omega^{a\phantom{a}}_{\phantom{a}c}\wedge \omega^{cb}=\Omega^{ab}=-\Omega^{ba}
\label{StR}
\end{equation}
where the $\Theta^a$ are the torsion 2-forms while the $\Omega^{ab}$ are the curvature 2-forms. From these equations follow the Bianchi identities
\begin{equation}
d\Theta^a+\omega^a_{\phantom{a}b}\wedge \Theta^b=\Omega^a_{\phantom{a}b} \wedge \theta^b
\label{BT}
\end{equation}
\begin{equation}
d\Omega^{ab}+\omega^a_{\phantom{a}c}\wedge \Omega^{cb}-\Omega^{ad}\wedge \omega_d^{\phantom{d}b}=0
\label{BR}
\end{equation}
obtained by differentiation of (\ref{StT}) and (\ref{StR}).\\

The Einstein action for pure gravity in dimension 4 reads \cite{thi:1979}
\begin{equation}
\fraci_{\gr}=\frac{1}{16\pi G} \int\frac{1}{2}\varepsilon_{abcd}\Omega^{ab}\wedge \theta^c \wedge \theta^d
\label{E4}
\end{equation}
where $G$ is the Newton gravitational constant.\\
To simplify the notations in the following we adopt the natural units where $c=1, \hbar=1$ and $4\pi G=1$. The action (\ref{E4}) reads then
\[
\fraci_{\gr}=\frac{1}{4} \int\frac{1}{2}\varepsilon_{abcd}\Omega^{ab}\wedge \theta^c \wedge \theta^d
\]
which generalizes in dimension $n=s+1$ as
\begin{equation}
\fraci_\gr=\frac{1}{4}\int \Omega^{ab}\wedge \theta^\ast_{ab}
\label{E}
\end{equation}
where we use the notation \cite{mdv-mad:1987}
\begin{equation}
\theta^\ast_{i_1\dots i_q}=\frac{1}{(n-q)!} \varepsilon_{i_1\dots i_q i_{q+1}\dots i_n} \theta^{i_{q+1}}\wedge \dots \wedge \theta^{i_n}
\label{tast}
\end{equation}
with $\varepsilon_{i_1\dots i_n}$ completely antisymmetric such that $\varepsilon_{0\> 1\dots s}=1$.\\
The vanishing of the variation of $\fraci_\gr$ given by (\ref{E}) with respect to infinitesimal variations $\delta\omega^{ab}$ of the $\omega^{ab}$ implies the vanishing of the torsion
\begin{equation}
\Theta^a=0,\ \ \ \forall a\in \{0,\dots,s\}
\label{T0}
\end{equation}
while the vanishing of the variation of $\fraci_\gr$ with respect to infinitesimal variations $\delta\theta^a$ of the $\theta^a$ implies
\begin{equation}
\Omega^{bc}\wedge \theta^\ast_{abc}=0,\> \> \forall a \in \{0,\dots,s\}
\label{EE}
\end{equation}
which reduces to the vacuum Einstein equations whenever the torsion vanishes. In other words in the Einstein-Cartan formalism, (i.e. the ``tetrad-connection " form), the vacuum Einstein equations are the combination of (\ref{EE}) and (\ref{T0}).\\
Let us come back to the 4-dimensional case. Then the complete Holst action reads 
\cite{hol:1996}
\begin{equation}
\fraci_H=\frac{1}{4}\int \Omega^{ab}_{\phantom{a}\wedge} (\frac{1}{2}\varepsilon_{abcd}\> \theta^c \wedge\theta^d-\frac{1}{\gamma} \theta_a\wedge \theta_b)
\label{H4}
\end{equation}
where the additional dimensionless constant $\gamma\not= 0$ is the Immirzi parameter \cite{imm:1997}. The vanishing of the variation of $\fraci_H$ with respect to infinitesimal variations $\delta\omega^{ab}$ of the $\omega^{ab}$ implies
\[
\frac{1}{2} \varepsilon_{abcd} \nabla (\theta^c\wedge \theta^d)-\frac{1}{\gamma}\nabla (\theta_a \wedge \theta_b)=0
\]
which with the Lorentzian signature ($\Rightarrow \ast\ast=-I$) implies 
\[
(\gamma^2+1)\nabla (\theta^a \wedge \theta^b)=0
\]
where $\nabla$ is the exterior covariant differential. Thus one has $\nabla(\theta^a\wedge \theta^b)=0$ whenever $\gamma^2+1\not=0$ which is equivalent to (\ref{T0}) i.e. to the vanishing of the torsion. Noticing the identity following from (\ref{BT})
\[
d(\Theta^a\wedge \theta_a)=\Theta^a \wedge \Theta_a-\Omega_{ab}\wedge \theta^a\wedge \theta^b
\]
one sees that in the Holst action (\ref{H4}) one may replace the Holst term $-\frac{1}{\gamma} \Omega^{ab}\wedge \theta_a\wedge \theta_b$ by $-\frac{1}{\gamma}\Theta^a \wedge \Theta_a$. Thus, apart from boundary terms, the Holst action can be put in the form
\begin{equation}
\fraci'_H=\frac{1}{4}\int \frac{1}{2}\varepsilon_{abcd} \Omega^{ab}\wedge \theta^c\wedge \theta^c-\frac{1}{\gamma} \Theta^a \wedge \Theta_a
\label{Hprime}
\end{equation}
with Holst term quadratic in the torsion. It follows then that the vanishing of the variation of $\fraci_H$ $\sim \fraci'_H$ with respect to infinitesimal variations $\delta\theta^a$ of the $\theta^a$ implies (\ref{EE}) whenever the $\Theta^a$ vanishes. In other words the Holst action $\fraci_H$ leads to the vacuum Einstein equations whenever $\gamma^2+1\not=0$ (in the case of Lorentzian signature). This is of course not a discovery and as well known similar properties hold in the case of Euclidean signature.

\section{Actions for the vacuum Einstein equations}
The discussion of the previous section for the Holst action is an illustration of the fact that if one adds to the Einstein action (\ref{E}) a term of order $\geq 2$ in the torsion which is such that the vanishing of the variation of the new action with respect to infinitesimal variations $\delta\omega^{ab}$ of the $\omega^{ab}$ implies the vanishing of the torsion then the new action leads to the vacuum Einstein equations. This is also the very reason for our construction in the following.\\
Let $x\mapsto F(x)$ be a smooth function on $\mathbb R$ vanishing at the origin, i.e. such that $F(0)=0$. Let us consider the following action
\begin{equation}
\fraci=\frac{1}{4} \int \Omega^{ab} \wedge \theta^\ast_{ab} + F ((\Theta,\Theta))\ \text{vol}
\label{AI}
\end{equation}
where the volume form vol is given by $1^\ast$ with the notation (\ref{tast}) and where $(\Theta, \Theta)$ is defined by 
\begin{equation}
\Theta^a\wedge \ast \Theta_a=(\Theta, \Theta)\ \text{vol}
\label{Scalp}
\end{equation}
with $\ast \Theta_a$ being the $(n-2)$-form Hodge-dual of the 2-form $\Theta_a$. That is
\[
\ast(\Theta_{abc} \theta^b \wedge \theta^c)=\Theta_a^{{\phantom a}bc} \theta^\ast_{bc}
\]
with the notation (\ref{tast}).\\
In view of the previous discussion, in order that $\fraci$ given by (\ref{AI}) leads to the vacuum Einstein equations, it is sufficient that the vanishing of the variation of $\fraci$ with respect to infinitesimal variations $\delta\omega^{ab}$ of the $\omega^{ab}$ implies the vanishing of the torsion (see Remark 1 below) i.e.
\begin{equation}
\frac{\delta\fraci}{\delta\omega^{ab}_{\phantom{ab}c}}=0\> (\forall a,b) \Rightarrow \Theta^c_{\phantom{c}ab}=0,\> (\forall c)
\label{ITO}
\end{equation}
where $\omega^{ab}=\omega^{ab}_{\phantom{ab}c} \theta^c$ and $\Theta^c=\Theta^c_{\phantom{c}ab}\theta^a\wedge \theta^c$.
Since both sides of (\ref{ITO}) have the same number of ``parameters"  the above implication will be generically an equivalence. Thus all that one has to do is to find the conditions that the function $F$ has to satisfy in order to have (\ref{ITO}). We shall show that there is an infinity of functions satisfying these conditions even among the polynomials.\\
Let us compute $\frac{\delta \fraci}{\delta\omega^{ab}_{\phantom{ab}c}}$ for $\fraci$ given by (\ref{AI}). Notice first that 
\begin{equation}
\frac{\partial F((\Theta,\Theta))}{\partial \omega^{ab}_{\phantom{ab}c}}=F'((\Theta,\Theta))((\theta^c\wedge \theta_b,\Theta_a)-(\theta^c\wedge \theta_a,\Theta_b))
\label{DF}
\end{equation}
where $F'(x)=\frac{dF}{dx}(x)$ denotes the derivative of $F$. It follows that the vanishing of  the $\frac{\delta\fraci}{\delta\omega^{ab}_{\phantom{ab}c}}$ reads
\begin{equation}
\Theta^c_{\phantom{c}ab}+\delta^c_{\phantom{c}a}\Theta^r_{\phantom{r}br}-\delta^c_{\phantom{c}b}\Theta^r_{\phantom{r}ar}+F'((\Theta,\Theta))(\Theta_{a\phantom{c} b}^{\phantom{a} c\phantom{b}}-\Theta_{b\phantom{c} a}^{\phantom{b} c\phantom{a}})=0
\label{DI0}
\end{equation}
Let us decompose $\Theta_{cab}$ into 3 disjoint representations of the Lorentz group $0(s,1)$ as
\begin{equation}
\Theta_{cab}=\Lambda_{cab}+(g_{ca}\Lambda_b-g_{cb}\Lambda_a)+T_{cab}
\label{Drep}
\end{equation}
where $\Lambda_{cab}$ is completely antisymmetric in $c,a,b$ and where $T_{cab}$ satisfies
\[
\left\{
\begin{array}{l}
T_{cab}+T_{cba}=0\\
g^{ca}T_{cab}=0\\
T_{cab}+T_{abc}+T_{bca}=0
\end{array}
\right.
\]
in other words $T$ has trace=0 and a vanishing completely antisymmetrized projection. The equation (\ref{DI0}) reads 
\begin{equation}
(\Fg'+1)T_{cab}+(\Fg'-s+1)(g_{ca}\Lambda_b-g_{cb}\Lambda_a)+(1-2\Fg')\Lambda_{cab}=0
\label{IDI}
\end{equation}
where we have set $\Fg'=F'((\Theta,\Theta))$.\\
It follows that if the image $\Im(F')$ of $F'$ is disjoint of the points $-1, \frac{1}{2}$ and $s-1$, i.e.
\begin{equation}
\Im(F')\cap \{-1,\frac{1}{2},s-1\}=\emptyset
\label{CT0}
\end{equation}
then $\frac{\delta \fraci}{\delta\omega}=0$ implies and in fact is equivalent to the vanishing of the torsion. Thus for any $F$ satisfying (\ref{CT0}) the action $\fraci$ given by (\ref{AI}) leads to the vacuum Einstein equations. It is clear that there are infinitely many such functions. In particular polynomials of the form
\[
a_1 x+\dots + a_{2p+1}x^{2p+1}
\]
will satisfy (\ref{CT0}) if the $a_\alpha$ are real such that
\[
a_{2p+1}>0\ \ \ \text{and}\ \ \ (2p+1)a_{2p+1}x^{2p}+\dots+a_1>s-1
\]
or 
\[
a_{2p+1}<0\ \ \ \text{and}\ \ \ (2p+1)a_{2p+1}x^{2p}+\dots+a_1<-1
\]
which just corresponds to a region of the parameters $a_\alpha$.\\

\noindent \underbar{Remarks}. \\
1 - Although more or less classical and easy to prove, the fact that $\frac{\delta \fraci}{\delta\omega^{ab}_{\phantom{ab}c}}=0$ is equivalent to the vanishing of the variation of $\fraci$ with respect to infinitesimal variations $\delta\omega^{ab}$of the $\omega^{ab}$ needs perhaps some explanations. Let $x^\lambda$ be some local coordinates and let $\omega^{ab}=\omega^{ab}_{\phantom{ab}\lambda} dx^\lambda$ be the components of the connection 1-form. One has 
\[
\omega^{ab}_{\phantom{ab}\lambda}=\omega^{ab}_{\phantom{ab}c} \theta^c_\lambda
\]
where the coframe $\theta^a$ reads $\theta^a=\theta^a_\lambda dx^\lambda$. It follows that
\[
\delta\fraci=\int \frac{\delta\fraci}{\delta\omega^{ab}_{\phantom{ab}\lambda}}\delta\omega^{ab}_{\phantom{ab}\lambda}+\frac{\delta\fraci}{\delta\theta^c_\rho}\delta\theta^c_\rho
= \int\frac{\delta\fraci}{\delta\omega^{ab}_{\phantom{ab}\lambda}}(\delta\omega^{ab}_{\phantom{ab}c}\theta^c_\lambda+\omega^{ab}_{\phantom{ab}c}\delta\theta^c_\lambda)+ \frac{\delta\fraci}{\delta\theta^c_\lambda}\delta\theta^c_\lambda
\]
and therefore one has
\[
\frac{\delta\fraci}{\delta\omega^{ab}_{\phantom{ab}c}}=\frac{\delta\fraci}{\delta\omega^{ab}_{\phantom{ab}\lambda}}\theta^c_\lambda
\]
which implies that $\frac{\delta\fraci}{\delta\omega^{ab}_{\phantom{ab}c}}=0$ is equivalent to $\frac{\delta\fraci}{\delta\omega^{ab}_{\phantom{ab}\lambda}}=0$ since $\det(\theta^c_\lambda)\not=0$.\\

\noindent 2 - Condition (\ref{CT0}) is not only sufficient but it is also necessary in order that the action $\fraci$ given by (\ref{AI}) leads to the vacuum Einstein equations. Otherwise, in view of (\ref{IDI}), the torsion $\Theta$ can take the form of one of the 3 terms of the decomposition (\ref{Drep}) with $(\Theta,\Theta)$ such that $F'((\Theta, \Theta))$ takes the value annihilating the corresponding coefficient in (\ref{IDI}) and thus $\fraci$ leads then to equations which admit other solutions than the solutions of the vacuum Einstein equations (although of course their generic solutions satisfy the vacuum Einstein equations).\\

\noindent 3 - At the beginning of this section we have pointed out the general fact that if one adds to the Einstein action (2.6) a term of order $\geq$ 2 in the torsion which is such that the vanishing of the variation of the new action with respect to infinitesimal variations $\delta\omega^{ab}$ of the $\omega^{ab}$ implies the vanishing of the torsion then the new action leads to the vacuum Einstein equations. The reason for this is that the variational principle for the $\delta\omega^{ab}$implies then by definition that the $\omega^{ab}$ are the component of the Levi-Civita connection while the variation of the new action with respect to the $\delta\theta^a$ just consists of the addition to the corresponding variation of the Einstein action (2.6) of a term of order $\geq 1$ in the torsion which therefore vanishes with the torsion (i.e. ``on shell"). Thus one sees that the condition of order $\geq 2$ in the torsion for the additional term is essential here (to get order $\geq 1$ for the variation). This condition is of course satisfied by $\frac{1}{4}\int F((\Theta,\Theta))$ vol with $x\mapsto F(x)$ smooth and such that $F(0)=0$.

\section{Adding the Holst term in dimension 4}

Let $F$ be as in Section 3 a smooth function on $\mathbb R$ with $F(0)=0$ and let $F'$ denote its derivative. Assume that we are in dimension 4 (i.e. $s=3$) and let us consider the action

\begin{equation}
\fraci_\gamma=\frac{1}{4}\int \Omega^{ab}\wedge (\frac{1}{2}\varepsilon_{abcd} \theta^c \wedge \theta^d-\frac{1}{\gamma}\theta_a\wedge \theta_b)+F ((\Theta, \Theta))\ \text{vol}
\label{AIH}
\end{equation}
that is the previous action (\ref{AI}) in dimension 4 with the addition of the Holst term. The equations
\[
\frac{\delta\fraci_\gamma}{\delta\omega^{ab}_{\phantom{ab}c}}=0
\]
reads then
\begin{eqnarray}
\Theta^c_{\phantom{c}ab}&+&\delta^c_{\phantom{c}a}\Theta^r_{\phantom{r}br}-\delta^c_{\phantom{c}b}\Theta^r_{\phantom{r}ar}+F'((\Theta,\Theta))(\Theta_{a\phantom{c} b}^{\phantom{a} c\phantom{b}}-\Theta_{b\phantom{c} a}^{\phantom{b} c\phantom{a}})\nonumber\\
&+&\frac{1}{2\gamma} \varepsilon^{ck\ell m}(\Theta_{ak\ell} g_{bm}-\Theta_{bk\ell}\  g_{am})=0
\label{DIH0}
\end{eqnarray}
and implies the vanishing of the torsion (i.e.$\Theta=0$) whenever $\Fg'=F'((\Theta,\Theta))$ is such that
\begin{equation}
(\Fg'+1)^2+\frac{1}{\gamma^2}\not= 0
\label{P1}
\end{equation}
and 
\begin{equation}
(\Fg'-\frac{1}{2})(\Fg'-2)+\frac{1}{\gamma^2}\not=0
\label{P2}
\end{equation}
from which follows that if $F$ is such that the image $\Im (F')$ of $F'$ is disjoint from the roots $-1\pm \frac{i}{\gamma}$ of the polynomial in $\Fg'$ on the left handside of (\ref{P1}) and from the roots $r_1$ and $r_2$ of the polynomial in $\Fg'$ on the left hand side of (\ref{P2}) then the action $\fraci_\gamma$ given by (\ref{AIH}) leads to the vacuum Einstein equations.\\
It is clear that there are infinitely many such functions $F$ (with $F$ smooth $F(0)=0$) whenever $\gamma^2+1\not=0$. In particular for $\gamma$ real, polynomials of the form
\[
a_1x +\dots +a_{2p+1}x^{2p+1}
\]
will satisfy (\ref{P1}) and (\ref{P2})if the $a_\alpha$ are real such that 
\[
a_{2p+1}>0\hspace{0,5cm} \text{and}\hspace{0,5cm}  (2p+1) a_{2p+1} x^{2p}+\dots + a_1> \sup (r_1, r_2)
\]
or 
\[
a_{2p+1}<0\hspace{0,5cm} \text{and}\hspace{0,5cm}\ (2p+1) a_{2p+1} x^{2p}+\dots + a_1< \inf (r_1, r_2)
\]
whenever $r_1$ and $r_2$ are reals. Otherwise, if all the roots have a nonvanihing imaginary part, any real valued function $F$ is good.\\

If (\ref{P1}) or (\ref{P2}) is not identically satisfied then the equations corresponding to $\fraci_\gamma$ admit other solutions that the vacuum Einstein equations. In fact, using the invariant decomposition (\ref{Drep}), if (\ref{P1}) is not satisfied one has solutions with torsion of the form
\[
\Theta_{abc}=T_{abc}
\]
 while if (\ref{P2}) is not satisfied one has solutions with torsion of the form
\[
\Theta_{abc}=\Lambda_{abc}+g_{ab}\Lambda_c-g_{ac} \Lambda_b
\]
which corresponds to Remark 2 for the case $\gamma^2=\infty$.\\

Notice that setting $F=0$ one recovers the condition $\gamma^2+1\not=0$ of Section 2 while by letting $\gamma^2\rightarrow \infty$ one recovers the conditions of Section 3 for $s=3$ since $\lim_{\gamma^2\rightarrow \infty}(-1\pm \frac{i}{\gamma})=-1$, $\lim_{\gamma^2\rightarrow \infty} (r_1)=\frac{1}{2}$ and $\lim_{\gamma^2\rightarrow \infty} (r_2)=2 (=s-1)$.

\section{Conclusion}

We have shown that there is an infinity of local actions obtained by adding to the Einstein action torsion terms of the form $\int F((\Theta,\Theta)) \text{vol}$ (with $F(0)=0$) which lead to the vacuum Einstein equations. We have shown that this is still true in dimension 4 if one adds the Holst term. In arbitrary dimensions $s+1$, there are analogs of the Holst term and it is clear that the same constructions work when one adds these terms ; only the critical values that $F'$ must avoid change. The same holds obviously for the Riemannian and the other signatures instead of the Lorentzian one discussed here.  The addition of a cosmological constant does not modify the picture.
\\

In presence of matter fields which couple to the connection these actions lead to inequivalent theories. This gives thefore some additional flexibility for adapting the theory to eventual future observations.\\

These actions should also lead to different quantum theories (when defined) even without matter. This is suggested in particular by the path integral formulation of quantum theory. This would also appear in the Hamiltonian formalism since it is clear that the canonical variables are sensitive to the additional terms and thus canonical quantization will lead to different algebras in terms of the fields $\theta^a,\omega^{ab}$.\\
\setcounter{footnote}{0}
Concerning the canonical formalism the following two important issues are worth noticing \footnote{We thank the referee for suggesting to discuss theses points}.\\
Firstly, in a perturbative approach around the Minkowski space the propagators of the linearized theory which are constructed with the canonical variables are sensitive to the additional terms and therefore the same is true for the UV perturbative behaviour of the theory. One could expect that increasing the order in the torsion of the additional term is UV regularizing. However, since one wishes background independence for the theory, the role of perturbative expansions is unclear and that kind of argument must be taken with care.\\
Secondly, the very reason for the introduction of the Holst term with the Immirzi parameter is to have canonical variables of Ashtekar type. From this point of view, it is very unlikely that the addition of the new terms here preserves this property but there are probably several roads to the quantization of space-time.\\
In any case, a careful analysis of the canonical formalism for the new actions proposed here is out of the scope of this article but deserves further attention.

\newpage
\bibliographystyle{plain}

\end{document}